\documentclass{aa}  
\usepackage{graphicx}
\usepackage{txfonts}
\def\ltsim{\hbox{\raise 2pt \hbox {$<$} \kern-1.1em \lower 4pt \hbox {$\sim$}}}
\def\ltapprox{\hbox{\raise 2pt \hbox {$<$} \kern-1.1em \lower 5pt \hbox
{$\approx$}}}
\def\gtsim{\hbox{\raise 2pt \hbox {$>$} \kern-1.1em \lower 4pt \hbox {$\sim$}}}
\def\gtapprox{\hbox{\raise 2pt \hbox {$>$} \kern-1.1em \lower 5pt \hbox
{$\approx$}}}

\def\eps{\varepsilon}
\def\cre{{\rm CRe}}

\def\degrees{$^{\circ}$}
\begin{document}
   \title{Low-Frequency study of two clusters of galaxies:\\ A2744 and A2219}

   \subtitle{}

   \author{E. Orr{\`u} \inst{1, 2}  
           \and  
           M. Murgia \inst{2,3} 
           \and 
           L. Feretti \inst{3} 
           \and
           F. Govoni \inst{2} 
           \and
           G. Brunetti \inst {3} 
           \and 
           G. Giovannini \inst{3,4} 
           \and 
           M. Girardi \inst{5,6} 
           \and 
           G. Setti \inst{3,4}
          }

   \offprints{E. Orr{\`u} eorru\_s@ira.inaf.it}

   \institute{
     Dipartimento di Fisica, Universit{\`a} degli Studi di Cagliari, 
     Cittadella Universitaria, I-09042 Monserrato (CA), Italy
     \and
     INAF\,-\,Osservatorio Astronomico di Cagliari, Loc. Poggio dei Pini, Strada 54,
     I-09012 Capoterra (CA), Italy
     \and
     INAF\,-\,Istituto di Radioastronomia, Via Gobetti 101, I-40129 Bologna, Italy
     \and
     Dipartimento di Astronomia Universit{\`a} degli Studi di Bologna, 
     Via Ranzani 1, 40127 Bologna, Italy
     \and
     Dipartimento di Astronomia, Universit{\`a} degli Studi di Trieste, 
     via Tiepolo 11, 34131 Trieste, Italy 
     \and
     INAF\,-\,Osservatorio Astronomico di Trieste, 
     via Tiepolo 11, 34131 Trieste, Italy 
     }

   \date{Received; accepted}

 \abstract
   {}
   { 
Spectral index images can be used to constraint the energy spectrum of relativistic electrons and magnetic field distribution 
in radio halos and relics, providing useful information to understand their formation, evolution and connection to cluster merger
 processes.
}
   {
We present low-frequency images of the two clusters of galaxies:  
A2744 and A2219, in which a wide diffuse emission is detected. 
Observations were made with the Very Large Array at the frequency of 325\,MHz. 
For both clusters deep Very Large Array 1.4\,GHz observations are available. 
Combining the 325\,MHz and 1.4\,GHz data, we obtained the spectral index images 
 and the brightness radial profiles of the diffuse radio emission with a resolution 
of $\sim$ 1\arcmin~.
}
   { The azimuthally averaged spectral index in A2744 is constant to a value close to $\alpha\simeq 1$ up to a distance of 1 Mpc from the 
cluster center. However, the spectral index image shows the presence of localized regions in which the radio spectrum is significantly
 different from the average. The observed spectral index variations range from a minimum of $\alpha \simeq 0.7\pm0.1$ to a maximum $\alpha \simeq 1.5\pm0.2$. 
From the comparison of the spectral index with the X-rays data it is found 
for the first time that the flat spectrum regions of the radio halo tend to have higher temperature.
 In the case of  A2219, the radio emission in the central regions of the cluster is dominated by the blend of discrete sources.  The azimuthally averaged 
radio spectrum is $\alpha\simeq 0.8$ in the central region of the cluster and is close to a value of $\alpha\simeq 1$ in the radio halo. The limited sensitivity of the 325\,MHz image does not allowed us to detect all the radio halo structure seen at 1.4\,GHz and therefore no constrains on the
 point-to-point variations of the spectral index have been obtained for this cluster.}
   {}


 \maketitle
%
%____________________________________________________

\section{Introduction}

In the hierarchical merging scenario, large-scale-structures form as the result of 
several merger events (e.g. Evrard \& Gioia 2002).
Clusters of galaxies are the most massive gravitationally bound objects in the Universe and 
they are structures still forming at the present epoch.

The presence of wide diffuse radio sources associated with the intra-cluster medium (ICM) has been detected
in an increasing number of massive, merging, clusters of galaxies.
These synchrotron  diffuse radio sources are characterized by a typical size of about 1 Mpc,  
low surface brightness ($\simeq~10^{-6}$ Jy/arcsec$^2$ at 1.4 GHz) and steep radio 
spectrum\footnote{$S(\nu)~\propto~\nu^{-\alpha}$}
($\alpha\geq$1). They are  classified as halos, if they are located in the cluster center, or relics, 
if they are in the peripheral regions of the cluster (e.g. Giovannini \& Feretti 2002).
Radio halos and relics demonstrate the existence of relativistic 
electrons and large scale magnetic fields in the ICM. 
Many efforts have been done to explain the origin of halos and relics.
The relic emission is believed to be caused by the propagation of shock waves 
produced during cluster merger events. In this case, the radio emission traces the rim of the 
shock wave in which electrons are injected and/or re-accelerated 
(e.g. Rottgering et al. 1997; En{\ss}lin et al.  1998). 
In the case of radio halos, it is required that either the electrons are 
re-accelerated ({\it primary models}: e.g. Tribble 1993, Brunetti et al. 2001; Petrosian 2001;
Brunetti et al. 2004; Fujita et al. 2003; Cassano \& Brunetti 2005)
or continuously injected over the entire cluster volume by hadronic collisions 
({\it secondary models}: e.g. Dennison 1980; Blasi \& Colafrancesco 1999; 
Dolag \&  En{\ss}lin 2000).
On the other hand, the strength and structure of the magnetic fields have been estimated 
and simulated with different methods (see e.g. Carilli \& Taylor 2002 and 
Govoni \& Feretti 2004 for reviews).     
Spectral index maps of radio halos and relics are promising tools in the understanding of their 
formation, evolution and connection to cluster merger processes. Moreover they provide important 
information on the energy spectrum of relativistic electrons and magnetic field distribution 
(Brunetti et al.  2001, Feretti et al. 2004).

Given the large angular sizes ($\sim 10 \arcmin$) and steep radio spectrum of halos and relics, 
suitable spectral index images can only be obtained at frequencies lower than $\sim$ 1 GHz.
Nowadays Coma (Giovannini et al.  1993), A2163, A665 (Feretti et al. 2004) and 
A3562 (Giacintucci et al.  2005) are the only radio halos for which spectral index 
maps were presented in the literature.
In line with primary models, in all these clusters the spectral index maps reveal patchy structure 
with, in some cases, a trend showing a
progressive steepening of the spectrum with increasing distance from the cluster 
center to the edge. 

\begin{figure*}[t]
\begin{center}
\includegraphics[width=18cm]{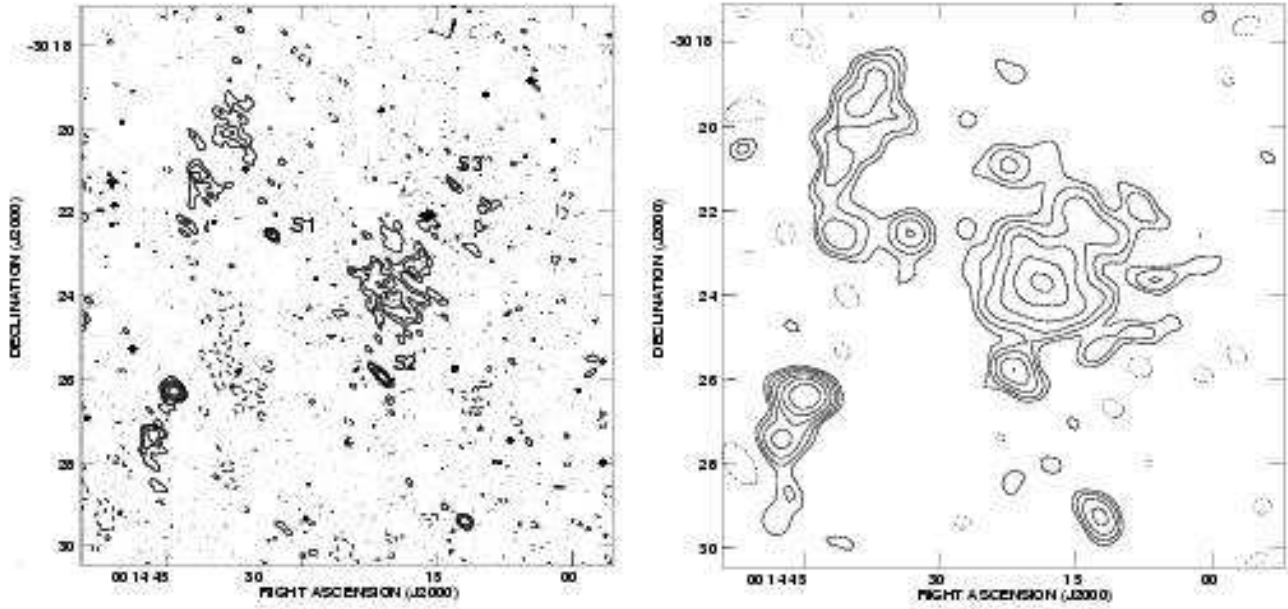}
\end{center}
\caption[] {A2744. Left panel:  Radio contours of the 325\,MHz  image obtained with the VLA in BnA configuration 
 are overlaid to the Digitized Sky Surveys optical image of the cluster.
The radio image has a resolution of 21\arcsec $\times$ 11\arcsec with a PA= 51\degr. The noise level is 0.9 mJy/beam.
Contours are scaled by $\sqrt{2}$ and the first two levels are -2 and 2 mJy/beam.
Right panel: Radio contours of the image at 325\,MHz obtained with the VLA in CnB 
configuration. The resolution is 56\arcsec $\times$ 44\arcsec with a PA= 58\degr. 
The noise level is 2.3 mJy/beam. Contours are scaled by $\sqrt{2}$ and the first two levels are -4.6 and 4.6 mJy/beam.}
\label{A2744cntr}
\end{figure*} 

In this paper we present 325\,MHz images of two clusters: 
Abell 2744 which presents a central radio halo and a peripheral relic
and Abell 2219 which hosts a radio halo. The general characteristics 
of A2744 and A2219 are listed in Table~\ref{clusters}.

The paper is organized as follows. In Sect. 2 we discuss the details about 
the radio observations and the data
reduction. In Sect. 3 we present the radio properties of A2744, we show data at 325\,MHz,
the spectral index map and profile obtained by combining the image at 325\,MHz 
with available data at 1.4\,GHz. Moreover we compare the
spectral index map with the cluster properties in the X-ray and optical wavelength.
In Sect. 4 the radio properties of A2219 and the new image at 325\,MHz are presented and 
the spectral index distribution obtained between 325\,MHz and 1.4\,GHz is shown and discussed.
The spectral index behavior in A2744
is analyzed in the framework of a particle re-acceleration model in Sect. 5.
Finally, in Sect. 6 we summarize the results of this paper.

Throughout this paper we adopt: 
$ H_{0} $ =71 km s$^{-1}$ Mpc$^{-1}$ , $\Omega_{\Lambda}$ = 0.73 and $\Omega_{m}$=0.27 . 

\begin{table}
\caption{Clusters properties.}
\begin{center}
\begin{tabular}{ccccc}
\noalign{\smallskip}
\hline                  
\noalign{\smallskip}    
Name   & $\alpha$(J2000) & $\delta$(J2000) & z & kpc$\slash\arcsec$   \\
       & ($h$ $m$ $s$)  & (\degr\ \arcmin\ \arcsec) &   &                \\               
\\ 
\hline
\hline
A2744 & 00:14:19  & $-$30:22:19  &  0.3080 &4.502 \\
A2219  &16:40:21  & $+$46:41:16  &  0.2256 &3.587 \\
\hline
\hline
\noalign{\smallskip}
\label{clusters}
\end{tabular}
\end{center}
\begin{list}{}{}
\item[] 
Col. 1: cluster name; 
Col. 2: and 3: cluster coordinates from NASA/IPAC extragalactic database (NED) ; 
Col. 4: redshift (Struble \& Rood 1999); 
Col. 5: arcsec to kpc conversion. 
\end{list}
\end{table}

\section{Observations and data reduction}

\begin{table*}
\caption{Summary of radio observations.}
\begin{center}
\begin{tabular}{cccccccc}
\noalign{\smallskip}
\hline                  
\noalign{\smallskip}    
Name   &$\alpha$(J2000) & $\delta$(J2000)  &${\nu}$ & $\Delta\nu$ & Configuration & Date & Duration \\
       & ($h$ $m$ $s$)  & (\degr\ \arcmin\ \arcsec)& MHz  & MHz&        & &hours            \\ 
\hline
\hline
A2744 &00 14 15 & $-$30 22 60 & 321.5, 327.5 & 3.125  & BnA&  2004-oct-04 & 2.5  \\
      &         &             & 321.5, 327.5 &  3.125 & BnA & 2004-oct-07 & 3.5 \\
      &         &             & 321.5, 327.5 &  3.125  & CnB & 2004-feb-06 & 3.5\\
      &         &             & 321.5, 327.5 &  3.125 & CnB & 2004-feb-12 & 3.5 \\
A2219 &16 40 15 & $+$46 41 60 & 321.5, 327.5 &  3.125  & B   & 2003-oct-18 & 2.5 \\
      &         &             & 321.5, 327.5  &  3.125  & C   & 2004-may-15 & 7.5 \\ 
\hline
\hline
\noalign{\smallskip}
\label{observation}
\end{tabular}
\end{center}
\begin{list}{}{}
\item[] 
Col. 1: cluster name;
Col. 2, 3: radio pointing position; 
Col. 4: observing frequency; 
Col. 5: bandwidth; 
Col. 6: VLA configuration; 
Col. 7: date of observations; 
Col. 8: time of integration.

\end{list}
\end{table*}

 Observations were conducted 
in the 327\,MHz band  with the  Very Large Array\footnote{The Very Large Array (VLA) is a facility of the National Radio Astronomy Observatory (NRAO). The NRAO is a facility of the National Science Foundation, operated under 
cooperative agreement by Associated Universities, Inc.} in different configurations. 
Observational parameters are summarized in Table~\ref{observation}.

A main problem in low-frequency observations are 
Radio Frequency Interferences (RFI) that corrupt the data.
Particularly in the 327\,MHz band, the internal electronic of the VLA 
gives rise to harmonics that are multiples of 5 and 12.5 MHz. 
To avoid this problem, a bandwidth of 3.125 MHz is used; since it is narrow it can be placed 
between the 
location characterized by the so called ``comb'' of RFI (Kassim et al. 1993).
The observations are usually made in spectral line mode,  dividing the bandwidth in several 
channels. This also reduces the bandwidth smearing, which is very strong at low frequency.

Data were calibrated and reduced with Astronomical Image 
Processing System (AIPS). We made the amplitude and bandpass calibration 
with the sources 3C48 and 3C286 respectively for A2744 and A2219. 
We used the sources 0025-260 and 1710+460 for the initial phase calibration for 
A2744 and A2219, respectively. 

A careful data editing has been made in order to excise RFI 
channel by channel. At the end of this procedure 16$\%$ of data were flagged in A2744. 
The data-set of the 2004-oct-4 in BnA configuration was too noisy and it is not 
presented here. For A2219 flagged data were 10$\%$.

In the final imaging, for A2744 the data were averaged to 7 channels 390\,kHz large, while 
for A2219 they were averaged to 6 channels 488\,kHz large. The data were mapped using a 
wide-field imaging technique, which corrects for distortions in the 
image caused by the non-coplanarity of the VLA over a wide field of view.
A set of small overlapping maps was used to cover the central 
area of about $\sim 2\degr$ in radius (Cornwell \& Perley 1992).
However, at this frequency confusion lobes of sources  far from the 
center of the field are still present. Thus, we also obtained images of strong sources 
in an area of about $\sim 60\degr$ in radius, searched in the NRAO VLA Sky Survey (NVSS, Condon et al. 1998) catalog. 
All these ``facets'' were included in the CLEAN and used for several loops of phase self-calibration (Perley 1999). The 
central frequency of the final images is 325\,MHz for both A2744 and A2219.

We corrected the final image for the primary beam effect.
For A2744 the achieved rms sensitivity is 0.9 mJy/beam in BnA configuration and 2.3 mJy/beam 
in CnB configuration. To improve the u-v coverage and sensitivity we combined the BnA and CnB data sets. The
 resulting image has a noise level of 1 mJy/beam. For A2219 the achieved sensitivity is 1.7 mJy/beam in both arrays. 
All these values are somewhat higher than the expected thermal noise levels,  because of the 
contribution of several factors: confusion, broad-band RFI, VLA generated RFI 
and some others still unknown.

\section{Properties of the cluster A2744}

A2744 hosts a central radio halo and a peripheral relic, detected in the NVSS 
by Giovannini et al. (1999) and confirmed in a deeper observation at 1.4\,GHz by
Govoni et al. (2001a).
The normalized radio and X-ray brightness profiles of the cluster
appear to be very similar (Govoni et al. 2001b), indicating that 
there could be an energetic relation 
between the X-ray thermal emitting gas and the relativistic radio emitting particles.

Girardi \& Mezzetti (2001) and Boschin at al. (2006), derived 
that galaxies are described by a non-Gaussian velocity distribution. 
They found two galaxy groups with a mass ratio of 3:1 which are separated by a 
line-of-sight velocity of $\Delta V\sim 4000$  km\,s$^{-1}$.
 The main one, the {\it low-velocity} group, has a velocity dispersion of $\sigma_{V}\simeq 1200-1300$ km\,s$^{-1}$.
The secondary one,  the {\it high-velocity} group, has a velocity dispersion of $\sigma_{V}\simeq 500-800$ km\,s$^{-1}$.

Another indication that the cluster is out of the equilibrium is given 
by Allen (1998) who found high  discrepancies between X-ray 
masses and lenses masses.

High resolution {\it Chandra} X-ray observations confirm the highly 
disturbed state of the cluster, indeed temperature and brightness gradients 
have been measured by Kempner \& David (2004). 
They found a main merger in the proximity of the cluster center and 
a smaller one in the North-Western region, where the presence of a sub-cluster is evident.
The peak of the radio halo is located near the cluster X-ray center but the radio halo emission
is spread up to the North-Western sub-cluster.

Unfortunately the field of view of {\it Chandra} does not include the
region where the relic is located.

\subsection{Low frequency image of A2744}

\begin{figure*}[t]
\begin{center}
\includegraphics[width=18cm]{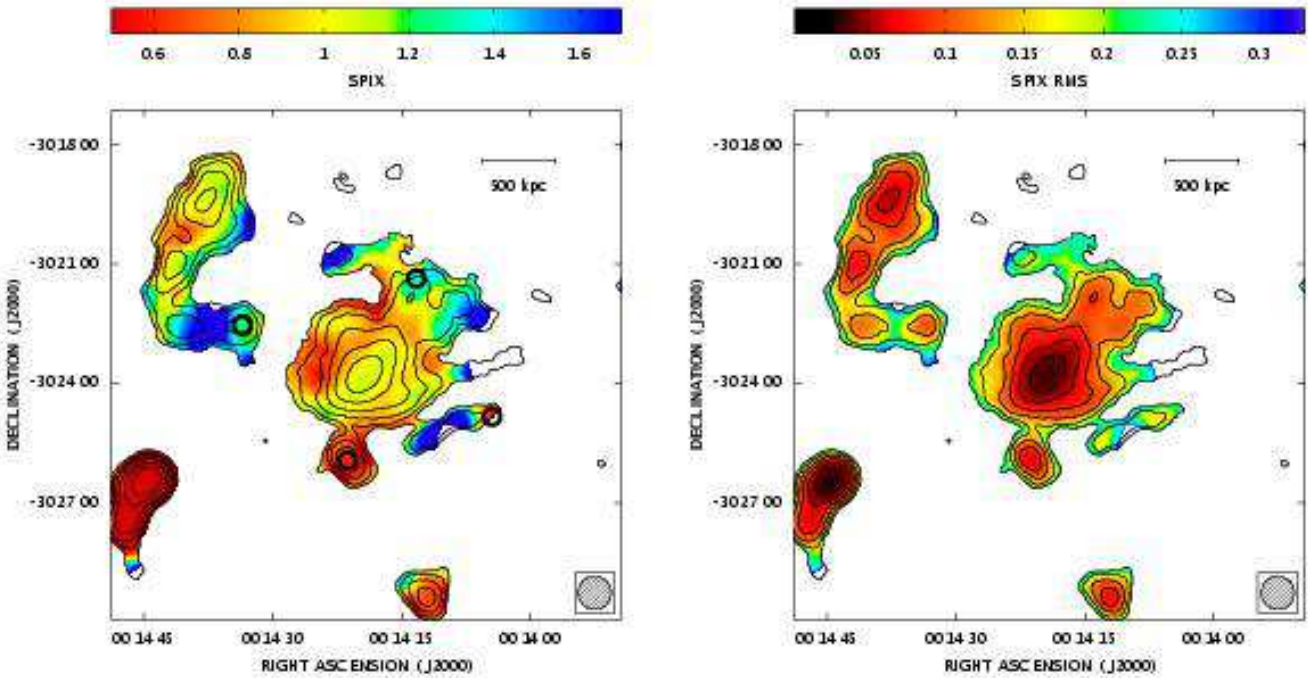}
\end{center}
\caption[] {A2744. Left panel:  The color-scale represents the spectral index image of A2744 between 325\,MHz and 1.4\,GHz, 
with a resolution of 50\arcsec$\times$50\arcsec. Pixels whose brightness was below 3$\sigma$ at 325\,MHz or 1.4\,GHz have been
 blanked. The cut is driven by the 325\,MHz image in most of the points. Contour levels are the radio image at 325\,MHz. Contours start at  3.9 mJy/beam (3$\sigma$) and scale by $\sqrt{2}$. The circles indicate the positions of discrete sources, the spectral index in these points is not representative of the radio halo emission. Right panel: The color-scale represents the image of the spectral index uncertainty.}
\label{A2744_spix}
\end{figure*}

\begin{figure*}[ht]
\begin{center}
\includegraphics[width=18cm]{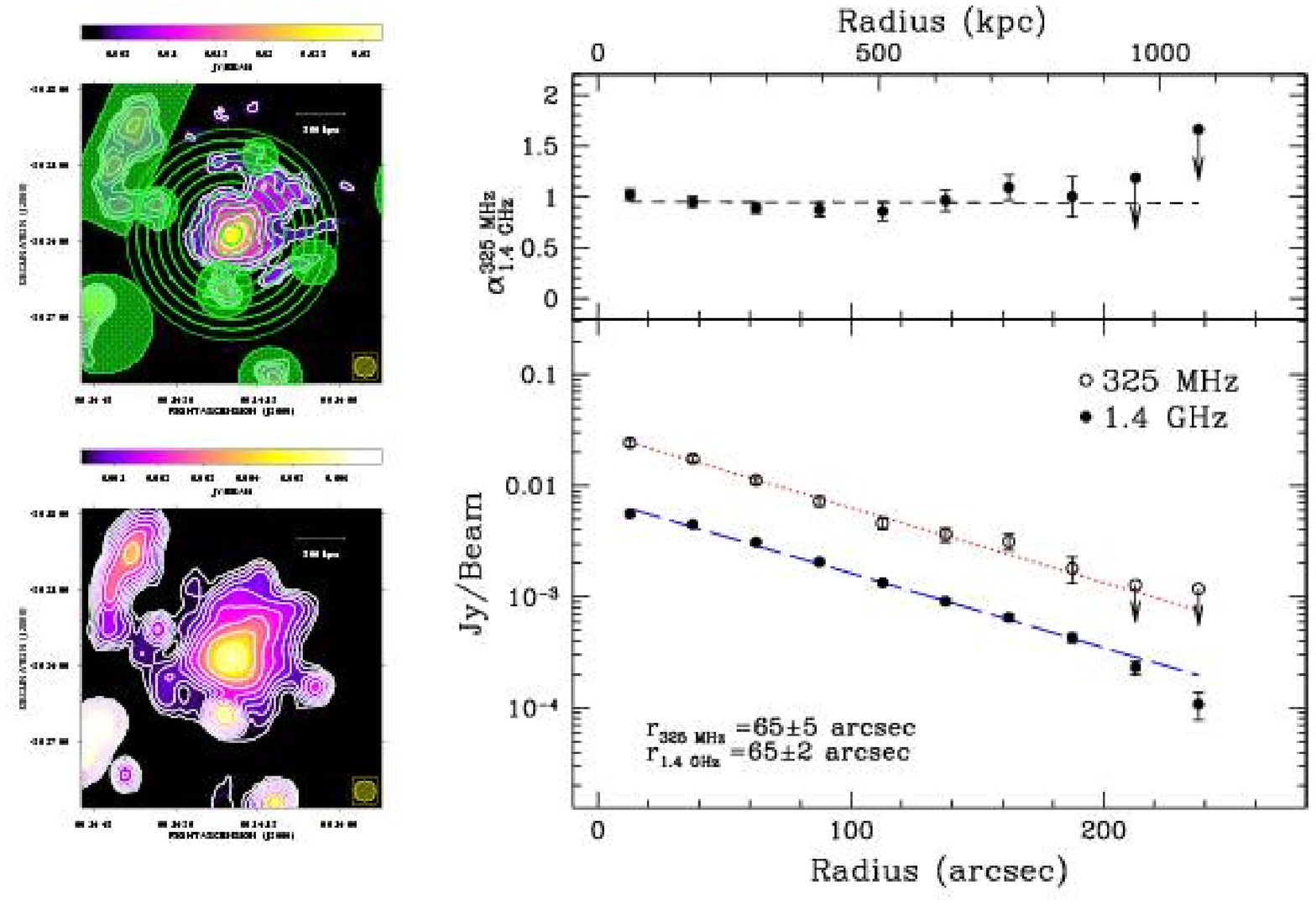}
\end{center}
\caption[] { A2744. Left: Colors represent the images at 325\,MHz (top panel) and 1.4\,GHz (bottom panel).
Contours represent the 325\,MHz and 1.4\,GHz images. Contours are spaced by $\sqrt{2}$ and start at 3.9 and 0.3  mJy/beam at 325\,MHz and 1.4\,GHz, respectively. The 
cross-hatched regions have been excluded from the statistics.
Right: Bottom panel shows the azimuthally averaged brightness profiles at 325\,MHz (open circles) and at 1.4\,GHz (solid circles). Arrows represent 
$3\sigma$ upper-limits. The lines represent the fit of an exponential profile to the data (see the text). In the top panel we plot the 
spectral index radial profile obtained from the brightness profiles.}  
\label{A2744_profi}
\end{figure*}

The BnA and CnB configuration images at 325\,MHz of A2744 are shown in Fig.~\ref{A2744cntr}.
In the BnA  image, which has a resolution of 21\arcsec $\times$ 11\arcsec, we detected both the 
halo and the relic. This confirms that the emission of the halo and the relic is not due to the 
blend of discrete sources. There are only three discrete sources in proximity of the radio halo visible in the 327\,MHz image, they are labeled with S1, S2 and S3 in the left panel of Fig.~\ref{A2744cntr}. 
Out of these sources, S2 could be associated with the member 
galaxy number 52 in Boschin at al. (2006), whereas S1 and S3 do not have any obvious optical
identification and are likely background objects. We note here that the strong discrete source detected
at RA=0$^{\rm h}$14$^{\rm m}$4$^{\rm s}$ and  DEC=$-30^{\rm d}$24$^{\rm m}$41$^{\rm s}$  in the 1.4 GHz image (Govoni 
et al. 2001a, see also Fig.\,\ref{A2744_profi} bottom left panel) is not detected in the 325\,MHz high 
resolution image of Fig.\,\ref{A2744cntr} (left) and is very faint in the 
lower resolution image of Fig.\,\ref{A2744cntr} (right). This is consistent with a source with a spectral index 
lower than $\alpha\,<$\,0.5.

The CnB configuration image is shown in the right panel of Fig.~\ref{A2744cntr}. This image has 
 a resolution of 56\arcsec $\times$ 44\arcsec. The combined BnA+CnB array image at 325\,MHz (not shown here), produced at 
intermediate angular resolution (26\arcsec $\times$ 16\arcsec), has a noise level of 1 mJy/beam, and confirms the
structures detected from the image of Fig.\,\ref{A2744cntr}.

The morphology of the diffuse emission is similar to that detected at 
1.4\,GHz by Govoni et al. (2001a) for both the halo and the relic, although 
 only the brightest regions of the halo can be seen at 325\,MHz due to 
 lower sensitivity of these observations with respect to that at 1.4\,GHz.
The radio halo in the center of the cluster has  a size of ${\sim}$ 6$^{\prime}$ ($\sim$1.6 Mpc).
The relic is located in the North-Eastern region of the cluster and it shows an elongation in North South direction
with a size of ${\sim}$ 6$^{\prime}\times 1^{\prime}$ ($\sim 1.6\times 0.3$ Mpc). 
Table \ref{result} summarizes the main physical parameters obtained from the 325\,MHz image
both for the halo and the relic.
The total flux of the halo, excluding discrete sources, is 218$\pm$10 mJy.
Using the total flux at 325\,MHz and 1.4\,GHz, measured in the same area, 
we calculated an average total spectral index of the halo of $\alpha$~$\sim$~1$\pm$0.1. 
The total flux of the relic is 98$\pm$7 mJy and its average spectral index is $\alpha$$\sim$1.1$\pm$0.1.
 
\begin{table}
\caption{Radio parameters.}
\begin{center}
\begin{tabular}{ccccc}
\noalign{\smallskip}
\hline                  
\noalign{\smallskip}    
Name   &type & $S_{325\,MHz}$ & $\alpha$  & $B_{eq}$ \\
       &     & mJy       &           & $\mu G$  \\ 
\hline
\hline
A2744  &H    &218$\pm$10  &1.0$\pm$0.1 &0.5  \\
       &R    &98$\pm$7  &1.1$\pm$0.1 &0.6  \\
A2219  &H    &232$\pm$17 &0.9$\pm$0.1 &0.4  \\
\hline
\hline
\noalign{\smallskip}
\label{result}
\end{tabular}
\end{center}
\begin{list}{}{}
\item[] 
Col. 1: cluster name;
Col. 2: source type (R=relic, H=halo);
Col. 3: flux density; the contribution of the point sources have been subtracted;
Col. 4: average spectral index between 325\,MHz and 1.4\,GHz;
Col. 5: equipartition magnetic field. 
\end{list}
\end{table}

\subsection{Spectral index analysis of A2744}
For the purposes of the spectral analysis we combined the BnA and CnB arrays at 325\,MHz and we compared this
 image with the C+D observation at 1.4\,GHz by Govoni et al. (2001). Images at both frequencies have been restored with
 the same beam of 50\arcsec$\times$50\arcsec. The final noise levels are 1.3 and 0.1 mJy/beam for the 325 and 1.4\,GHz images, respectively. 

The spectral index image of A2744, calculated between 325\,MHz and 1.4\,GHz, is shown in left panel of Fig.\,\ref{A2744_spix}. In the figure we
 indicate by circles the discrete sources, whose spectrum is not related to the radio halo. Right panel of Fig.\,\ref{A2744_spix}
shows the spectral index uncertainty, $\sigma_{\alpha}$. 
The spectral index image has been obtained by considering only those pixels where the brightness was above 3$\sigma$ level at both frequencies. 
Because of its higher noise, the cut is driven by the 325\,MHz image in most of the points. The very different sensitivities of the 325\,MHz and 1.4\,GHz images
 introduce a bias in the spectral index image. Indeed, the outermost source regions, whose brightness at 1.4\,GHz is at the lowest
 levels, can only be detected at 325\,MHz if their spectrum is steeper than $\sim 1.8$ (see discussion below).

\begin{figure*}[ht]
\begin{center}
\includegraphics[width=18cm]{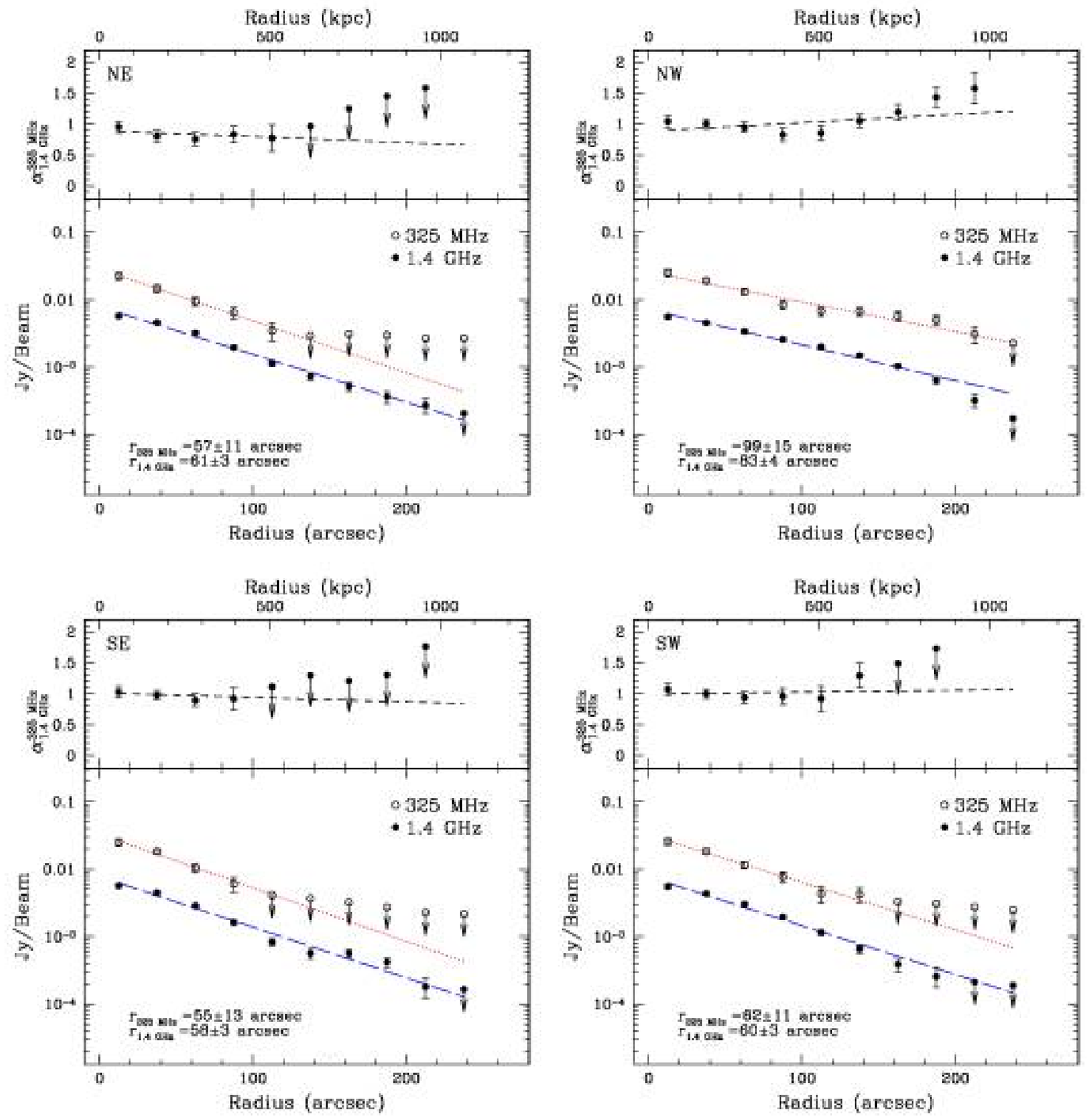}
\end{center}
\caption[] {A2744. Azimuthally averaged brightness profiles for four sectors. Arrows represent $3\sigma$ upper-limits. The lines represent the fit of an 
exponential profile to the data (see the text).}  
\label{A2744_sect}
\end{figure*}

The spectral index image is patchy. The central region has a spectral index of $\alpha \simeq 1.05\pm0.04$. A prominent region of flatter spectrum is present
 toward the East, with values of  $\alpha \simeq 0.7\pm0.1$. Another flat spectrum region is located North-West, before the asymmetric halo extension. Beyond this 
region, the spectrum is steep with values of the spectral index  $\alpha \simeq 1.5\pm0.2$. By comparison of the spectral index image with the error image (Fig.\,\ref{A2744_spix} left and right panel, respectively) it is evident that most spectral index features are statistically significant. 

In the relic, the spectral index trend shows a gradual steepening from values of $\alpha\sim$0.9~$\pm$~0.1 
to $\alpha\sim$1.6~$\pm$~0.2 from the outer to the inner rim. There are not drastic gradients along the main axis, 
as for the relics in A2256 (Clarke \& En{\ss}lin 2006) and A3667 (Rottgering et al. 1997).

We are interested to investigate if there is a systematic variation of the radio halo spectral index with radius as found in the radio halos of Coma (Giovannini et al. 1993), A665 and A2163 (Feretti el al 2004). 
The spectral index presented in  Fig.\,\ref{A2744_spix} is fully sampled to within a distance of about one core radius from the cluster center 
($r_{c}\simeq 115\arcsec\equiv 520$ kpc; Govoni et al. 2001). Since the 325\,MHz image has a noise level which is an order of magnitude higher than the 1.4\,GHz image,
 only the brightest regions of the radio halo at 325\,MHz can be detected beyond this distance. As stated above, the regions of the lowest 1.4\,GHz brightness are
 only detected if their spectrum is steeper than $\alpha>1.8$, otherwise their 325\,MHz emission is too faint to be detected at the sensitivity limit of the 325\,MHz 
 image. As a consequence, a radial steepening of the halo spectrum would be undetectable in the spectral index image, unless it is very strong, to values of $\alpha>1.8$ in the peripheral regions of the halo. Thus, to improve the statistics and take into account the limits on the 325\,MHz emission, we derived the radial 
trend of the spectral index from the azimuthally averaged brightness profiles at 1.4\,GHz and 325\,MHz without imposing any cut on the radio images.
In order to improve the statistics, we averaged the brightness in ten concentric annuli of $\sim$~25\arcsec~in width centered on the radio 
peak, as shown in top-left panel of Fig.~\ref{A2744_profi}.
The resulting azimuthally averaged brightness and spectral index profiles are shown in the right panel of Fig.~\ref{A2744_profi}. The errors associated with
 each point represent the $1 \sigma$ error on the average value while arrows represent $3\sigma$ upper-limits.

Brightness values below a level of 
 $3\sigma$ have been considered upper-limits.
We fitted the  brightness profiles with an exponential law of the form
\begin{equation}\label{expfit}
I(r)=I_0\,e^{-r/r_{\nu}}
\end{equation}
and we derived $r_{\nu}$, the e-folding radius of the brightness profile at both frequencies.
It results   $r_{325\,MHz}=65\arcsec\pm5\arcsec$ and $r_{1.4\,GHz}=65\arcsec~\pm2\arcsec$.
The e-folding radius at 325\,MHz is equal, to within the errors, to that at 1.4\,GHz, this implies that
 the azimuthally averaged spectral index is constant with the increasing distance from the
 cluster center as shown in the top plot of right panel of Fig.\,\ref{A2744_profi}. Here, the points  represent the spectral index obtained
 from the azimuthally average brightness profiles and the dashed line
 represents the trend expected from the fit of the exponential law. The azimuthally averaged spectral index in A2744 is constant to a 
value close to $\alpha\simeq 1$ up to a distance of 1 Mpc from the cluster center. This result implies that, although there are significant variations
 of the spectral index from point-to-point, on average the radio spectrum of the radio halo does not change with radius. 

\begin{figure}[h]
\begin{center}
\includegraphics[width=8cm]{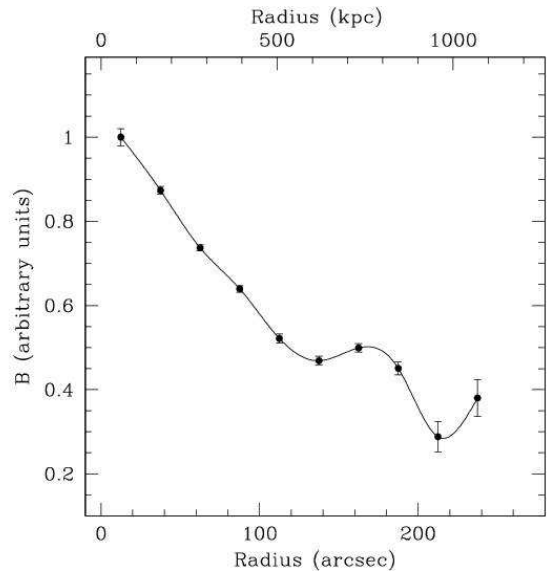}
\end{center}
\caption[] {Equipartition magnetic field radial profile obtained for the halo in A2744 for $\delta$=3. The profile has been normalized to its value at the 
center.}
\label{Bprofeq}
\end{figure}

In Fig.\,\ref{A2744_sect} we show
 the analysis of the azimuthally average brightness profiles for four cluster quadrants. We note that within the core radius ($r_{c}=115\arcsec$) $\alpha\simeq 1$ for all the 
four quadrants. However, at larger distances from the cluster center the four profiles differ. In agreement with the spectral index image, the steepest spectra are
 detected in the NW quadrant. Here, the spectral index reaches values of $\alpha\simeq 1.6 \pm 0.2$, after some flattening at about 80-120\arcsec. The flattest spectra
 are found in the SE sector, where the values of the outermost regions are constrained by upper limits.
\subsection{Equipartition magnetic field}
\label{equipartition}
Under the assumption that a radio source is in a minimum energy condition, 
it is possible to derive a zero-order estimate of the magnetic field 
strength averaged over the entire source volume.
In the classical equipartition assumption, 
considering a 
simplified shape of the spectrum of the emitting
electrons in the form:
\begin{equation}
N(\eps)d\eps=N_0\eps^{-\delta}d\eps
\end{equation}
the condition of minimum energy is obtained when 
the relativistic particle energy density $\eps_\cre=\int_{}^{}{\eps N(\eps)d\eps}$
is approximately equal to the magnetic field energy density $\eps_B = B^2/(8\pi)$.
In this assumption the magnetic field can be determined 
from the radio synchrotron luminosity and the source volume. 
The volume averaged magnetic field was evaluated within an ellipsoid
assuming a magnetic field entirely
filling the radio source, equal energy in relativistic 
protons and electrons, and a range of frequencies in which the
synchrotron luminosity is calculated from a low frequency cutoff of 10 MHz to
a high frequency cutoff of 10 GHz. For the spectral index of the electron energy spectrum 
of the halo we adopted the averaged emission spectral index
$\alpha\simeq 1.0$ which yields $\delta=3.0$. For the relic we use $\alpha\simeq 1.1$ which 
yields $\delta=3.2$. We estimated an equipartition magnetic field $B_{eq}\sim 0.5 ~\mu$G
in the halo and $B_{eq}\sim 0.6 ~\mu$G in the relic.

By assuming a low-frequency cut-off of 10\,MHz in the luminosity calculation is equivalent to
 assume a  low-energy cut-off of $\gamma_{min}\sim 2000$ in the particle energy spectrum. If alternatively  
we adopt a low-energy cut-off of $\gamma_{min}=100$ in the particle energy distribution
rather than a low-frequency cut-off in the emitted 
synchrotron spectrum (e.g. Brunetti, Setti \& Comastri 1997; 
Beck \& Krause 2005), we obtain B$'_{eq}\sim 1.0 ~\mu$G in 
the halo and B$'_{eq}\sim 1.3 ~\mu$G in the relic.

The radio synchrotron emissivity is given by:
\begin{equation}
 j_{\nu}\propto N_0\, B^{(\delta +1)/2}\, \nu ^{-(\delta-1)/2}.
\label{Jrela}
\end{equation}
Under equipartition conditions, if $\gamma_{min}$ is assumed constant 
with cluster radius, it is 
$N_0~\propto~B^2$ and therefore:
\begin{equation}
 j_{\nu}\propto B^{(\delta +5)/2}.
\end{equation}
 
According to this relation, assuming $\delta=3$, a spherical symmetry and using the
de-projected brightness profile at 325\,MHz, we obtain the 
equipartition magnetic field radial trend shown in Fig.~\ref{Bprofeq}. 
In each annulus, the de-projected profile has been obtained by subtracting the brightness 
contribution of the external shells. The magnetic field strength decreases by a factor of
 two from the cluster center to the halo periphery.

\subsection{Spectral index versus optical and X-ray bands.}

On the basis of optical and  X-ray data from the {\it Chandra} satellite, 
it has been proposed for A2744 (Girardi \& Mezzetti 2001; Kempner \& David 2004; 
Boschin et al. 2006) the following merger scenario:
a main merger has been observed, in the proximity of the cluster center, principally 
along the line of sight, in which two sub-clusters with mass ratio 3:1 and a non-zero 
impact parameter are interacting. A peripheral merger, with evidence of bow 
shock in the X-ray image, has been seen in the North-Western region. It involves a less
 massive sub-cluster and the central region in which the main merger is occurring.

Recent calculations (Cassano \& Brunetti 2005) show that a 
fraction $\sim$10\% of the thermal-cluster energy 
may be channeled in the form of turbulence in case of main 
(i.e., mass ratio $\leq$ 5:1) 
cluster mergers and this may power up giant radio halos 
such that observed in the case of A2744.
Although the minor mergers cannot directly generate giant radio halos, 
they may still power the particle acceleration process especially
if they happen in clusters which are already dynamically disturbed 
by previous mergers.

On the basis of the above merger scenario, 
we investigated if there is any connection between the
merger activity observed in the optical and X-ray bands and the spectral index distribution in the radio halo.

\begin{figure*}
\begin{center}
\includegraphics[width=18cm]{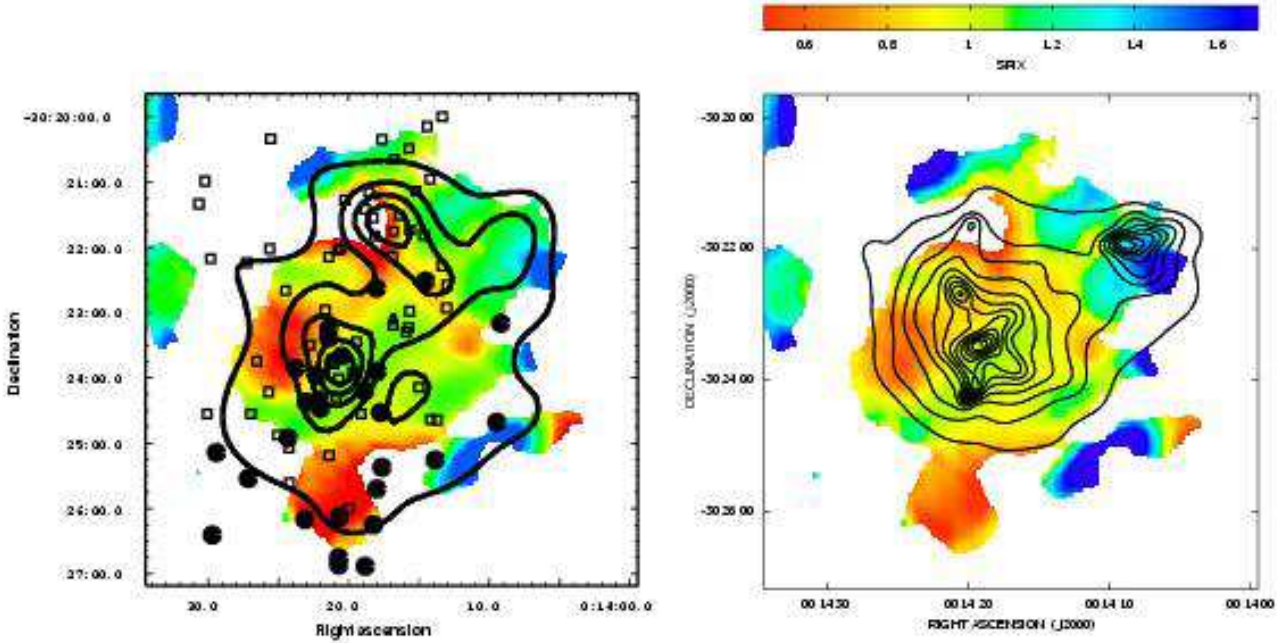}
\caption{  A2744: The spectral index images are shown in colors.  
Left panel: Contours are the spatial 2D distribution of likely members 
galaxies taken Boschin et al. (2006). Solid circles and open boxes represent the high-velocity and low-velocity galaxies, respectively. 
Right panel: Contours are the {\it Chandra} X-ray image, by Kempner \& David (2004).}
\label{2dpeaks}
\end{center}
\end{figure*}

The recent optical analysis by Boschin et al. (2006) based
on photometric and spectroscopic data of cluster galaxies
shows that A2744 is a very complex cluster from the optical point of view.
They pointed out a displacement between the peak of galaxy
distribution (see also the weak lensing analysis by Smail et al. 1997),
and the peak of X-ray emission,
as expected in the case of cluster mergers from numerical simulations
(e.g., Roettiger et al. 1997).
The peak of radio emission is displaced with 
respect to both optical and X-ray centers reinforcing the idea that
this cluster is dynamically very disturbed.

The two-dimensional distribution of likely members galaxies from 
Boschin et al. (2006) is shown in contours in the left panel of Fig.~\ref{2dpeaks}. Galaxies are
 concentrated in two regions: a  main clump, located at the center of the cluster, and a 
 secondary clump is placed at about 2.5$^{\prime}$ North with respect to the cluster center.
Spectroscopically these galaxies are characterized by two velocity groups separated by 
line-of-sight velocity of $\Delta V\sim 4000$  km\,s$^{-1}$.
The main one, the {\it low-velocity} group, has a velocity dispersion of $\sigma_{V}\simeq 1200-1300$ km\,s$^{-1}$. These
 galaxies are indicated with open boxes in left panel of Fig.~\ref{2dpeaks} and are distributed over the whole cluster.
The secondary one,  the {\it high-velocity} group, has a velocity dispersion of $\sigma_{V}\simeq 500-800$ km\,s$^{-1}$.
The galaxies of the  high-velocity group  are indicated with filled circles in left panel of Fig.~\ref{2dpeaks} and are
mainly concentrated in the South-West of the cluster. The high-velocity group  is likely merging with the main system
 and being responsible for the strongly disturbed central region. From the comparison with the spectral index image,
 there is no evident association between optical and spectral features. However, from the radial profiles presented in
 Fig.\,\ref{A2744_sect}, it is derived that the southern flat spectrum part of the radio halo coincides in projection
 with the high-velocity group of galaxies.

A spatial comparison between the spectral index image of A2744 and 
the {\it Chandra} X-ray brightness image (courtesy of J. C. Kempner, 
see Kempner \& David 2004) is shown in right panel of  Fig.\,\ref{2dpeaks}.
There is no evident correlation between the radio spectral index and the X-ray brightness features at the cluster center. We
 note, however, that the region of the NW group has a steep spectrum, while the stripe between the cluster center
 and the NW group, has a spectrum significantly flatter than the cluster center and the group itself. This is likely
 the region affected by the collision between the main cluster and the group.

\begin{figure*}[th]
\begin{center}
\includegraphics[width=15cm]{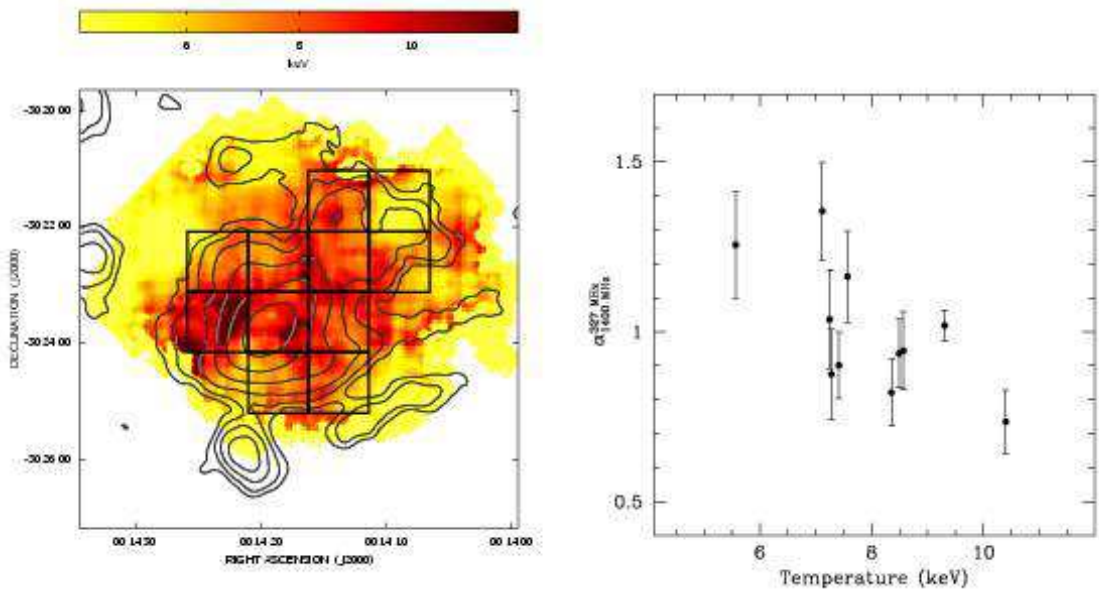}
\end{center}
\caption[] { Left panel: the grid used to calculate the spectral index-temperature scatter plot
 is overlayed to the {\it Chandra} temperature image (gray scale) by Kempner \& David (2004). The contours represent the 325\,MHz image. Right panel: Scatter plot of the radio  halo spectral index between 325\,MHz and 1.4\,GHz versus gas temperature.}
\label{spix-T}
\end{figure*}

We further compared the spectral index of the radio halo and the gas temperature image 
(kindly supplied by J. C. Kempner) of A2744. 
We averaged the spectral index and the temperature using a grid of rectangular boxes $63\arcsec\times63\arcsec$ in size 
 (see left panel of Fig.\,\ref{spix-T}), which is the extraction region of the temperature image (Kempner \& David 2004). 
The resulting scatter plot is shown in right panel of  Fig.\,\ref{spix-T}  where each point represents the average value 
in each cell of the grid. We found that the region with the highest gas temperature ($T\simeq 10$ keV) coincides with the flat spectrum clump in the  radio halo ($\alpha\simeq 0.7$) and that, in general, steep spectrum regions tend to be associated with lower temperature. This is the first time that such a correlation is detected in a radio
halo. 

\section{Properties of the cluster A2219}

A2219 hosts a giant radio halo detected in the NVSS 
by Giovannini et al. (1999) 
and confirmed in a deeper observation at 1.4\,GHz by Bacchi et al. (2003).
In the cluster center there are three strong radio sources identified as cluster 
galaxies by Owen et al. (1992).
Using {\it Chandra} archive X-ray data and optical spectra obtained with the {\it TNG}, 
Boschin et al. (2004) 
confirmed that the cluster is not dynamically relaxed. 
Indeed this cluster shows a SouthEast-NorthWest elongation, that is supported by the spatial 
distribution of the color-selected likely cluster members, the shape of the cD galaxy, 
the X-ray contours levels, the gradient in the velocity dispersion and the X-ray temperature.
From multi-wavelength analysis a very complex scenario results for the dynamic of the 
merger.      
Boschin et al. (2004) suggested that the cD galaxy in the center is suffering 
the consecutive merger of many clumps aligned in a filament for which 
the projection along the line of view is obliquely oriented.
They found a possible confirmation of this scenario in the elongated shape in 
the South East-North West direction observed in the radio contours map at 1.4\,GHz
by Bacchi et al. (2003). 

\subsection{Low frequency image of A2219}

\begin{figure*}[ht]
\begin{center}
\includegraphics[width=16cm]{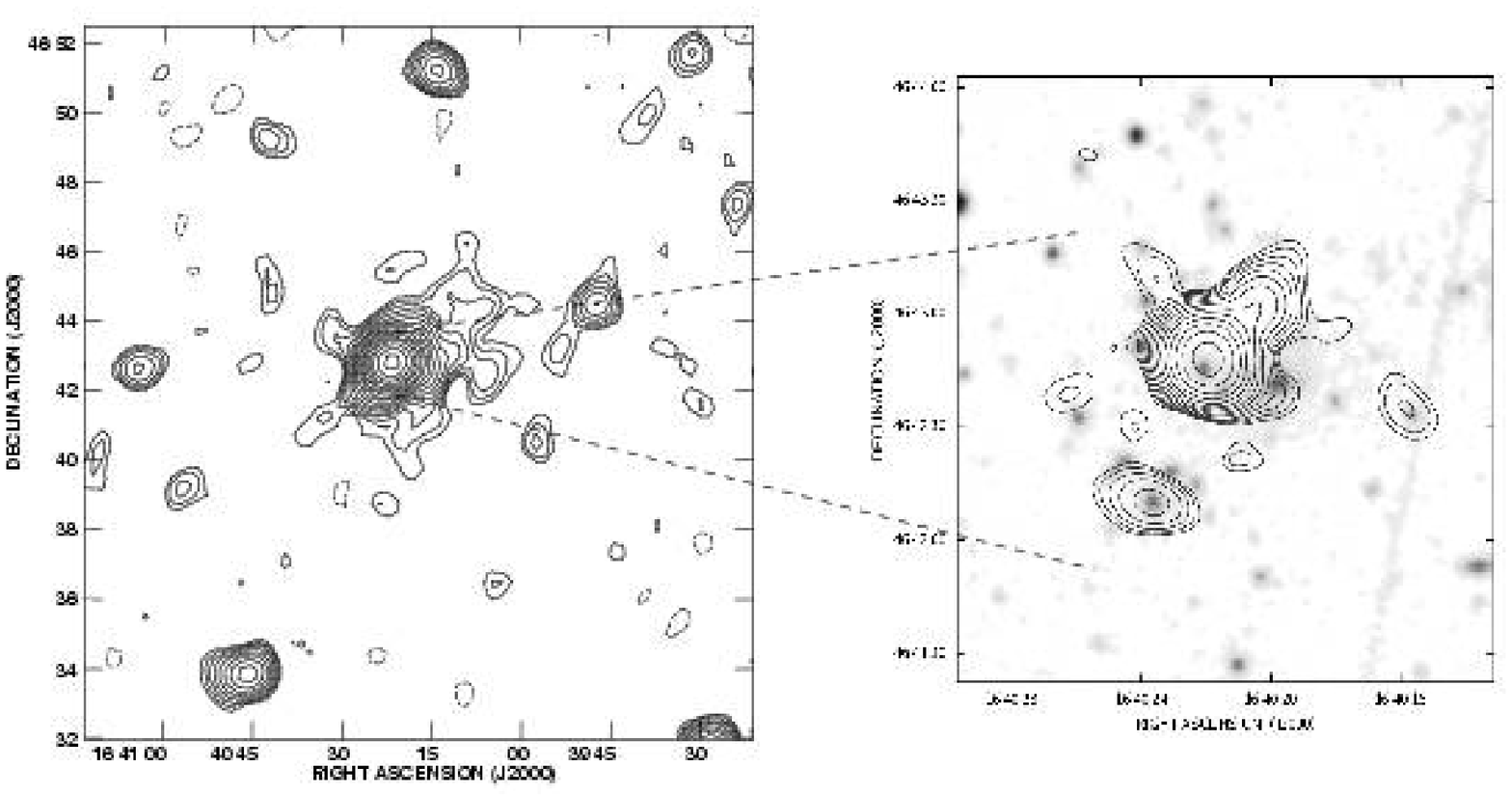}
\end{center}
\caption[] { A2219. Left panel: Radio contours of the image at 325\,MHz obtained with the VLA in C configuration. 
The resolution is $56\arcsec\times 53\arcsec$ with PA=$-85$\degrees.
The noise level is 1.7 mJy/beam, contours are scaled of $\sqrt{2}$ where the 
first two levels are $-3.4$ and 3.4 mJy/beam. Right panel: Radio contours of the image at 325\,MHz obtained 
with the VLA in B array overlaid onto the optical image (from Boschin et al. 2006). 
The resolution of the the radio image is $18\arcsec\times 16\arcsec$ with PA=41\degrees, 
the noise level is 1.7 mJy/beam, contours start at 7.5 mJy/beam (4.5$\sigma$ level) 
and are scaled of $\sqrt{2}$.}
\label{a2219cntr}
\end{figure*}

The A2219 diffuse radio emission at 325\,MHz obtained with the VLA in C configuration is shown in the image 
of Fig.~\ref{a2219cntr} (Left panel), which has a sensitivity level of 1.7 mJy/beam and 
a resolution of $56\arcsec\times 53\arcsec$. The obtained sensitivity  allows to have good detection 
of the central region of about $\sim $ 4.5\arcmin~ ($\sim$ 950 kpc) 
of the halo, in which the morphology is similar to that shown by Bacchi et al. (2003) with a 
NorthWest-SouthEast elongation.
The sensitivity of our observation is not enough to detect more extended regions 
as seen in low resolution image at 1.4\,GHz in Bacchi et al. (2003).

In the Right panel of Fig.~\ref{a2219cntr} we show a zoom of the central region of A2219, obtained 
with the VLA in B configuration, overlaid on the R-band optical image (Boschin et al. 2006). 
The sensitivity level is  1.7 mJy/beam, the resolution is $18\arcsec\times 16\arcsec$.  
It is evident that the radio emission at the cluster center is the blend of three cluster radio galaxies. 

The total flux, after the subtraction of discrete sources is 232$\pm$17 mJy and the average spectral 
index of the halo, calculated between 325\,MHz and 1.4\,GHz, is  $\alpha \sim$ 0.9$\pm$0.1.

The equipartition magnetic field, calculated within a sphere of 330\arcsec~ of radius following 
the classical approach, is $B_{eq}\sim 0.4~\mu$G (for $\delta=2.8$) and $B^{'}_{eq}\sim 0.7 ~\mu$G (assuming $\gamma_{min}=100$) with 
the approach given in Brunetti et al. (1997).

The main physical parameters of the radio halo in A2219 are summarized in Tab.~\ref{result}.

\subsection{Spectral index analysis of A2219}

\begin{figure*}[t]
\begin{center}
\includegraphics[width=18cm]{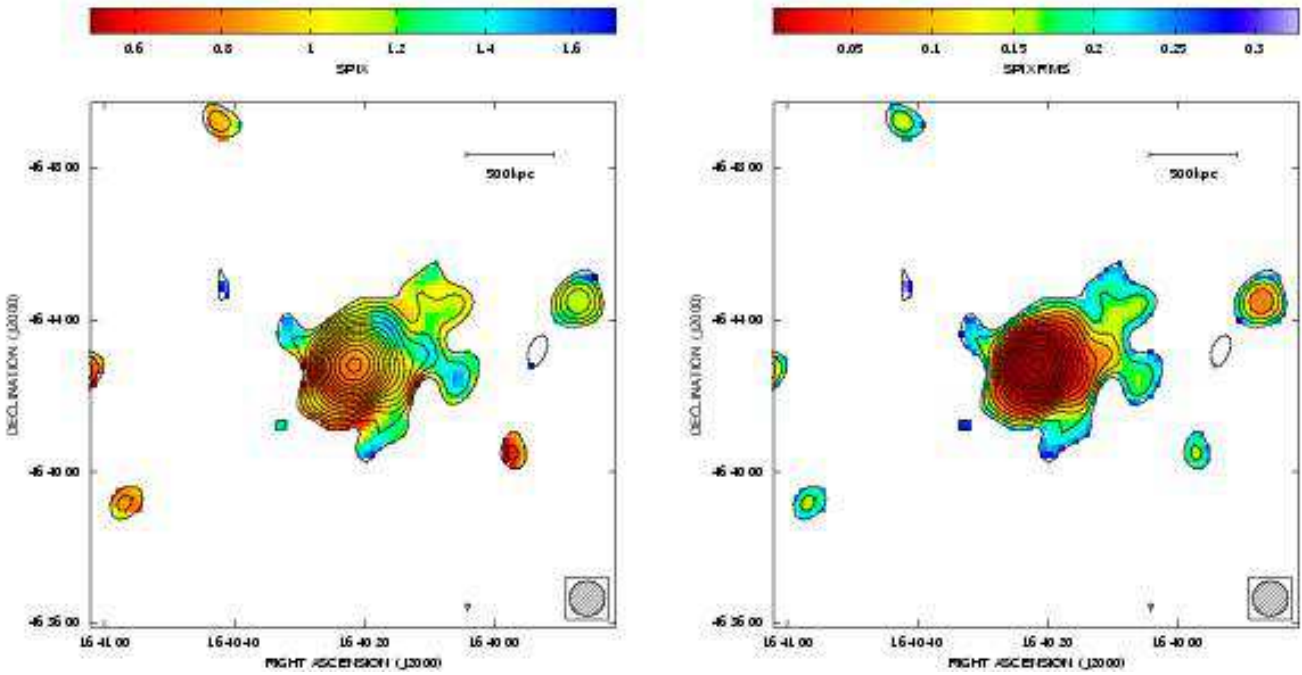}
\end{center}
\caption[] {A2219. Left panel:  The color-scale represents the spectral index image of A2219 between 325\,MHz and 1.4\,GHz, 
with a resolution of 56\arcsec$\times$56\arcsec. Pixels whose brightness was below 3$\sigma$ at 325\,MHz or 1.4\,GHz have been
 blanked. The cut is driven by the 325\,MHz image in most of the points. Contour levels are the radio image at 325\,MHz. Contours start at  5.1 mJy/beam (3$\sigma$) and scale by $\sqrt{2}$. Right Panel: The color-scale represents the image of the spectral index uncertainty.}   
\label{A2219_spix}
\end{figure*}

The spectral index image of A2219, with a resolution of $56\arcsec\times 56\arcsec$ 
is shown on left panel of Fig. \ref{A2219_spix}. The spectral index image, calculated between 325\,MHz and 1.4\,GHz,
have been made using images with the same restored beam, with the same pixel size and 
 considering only pixels with brightness values above 3$\sigma$ at both frequencies. Right panel of Fig.\,\ref{A2219_spix}
shows the $1-\sigma$ spectral index uncertainty.

The spectral index is $\alpha~\sim$ 0.8$\pm$0.05 in the inner 300\,kpc, where three blended compact sources are present.
 At larger distances, the spectral index can be considered representative of the halo emission and ranges between 
  $\alpha=1.0$ and $\alpha=1.5$ with a typical uncertainty of 0.2.

\begin{figure*}[t]
\begin{center}
\includegraphics[width=18cm]{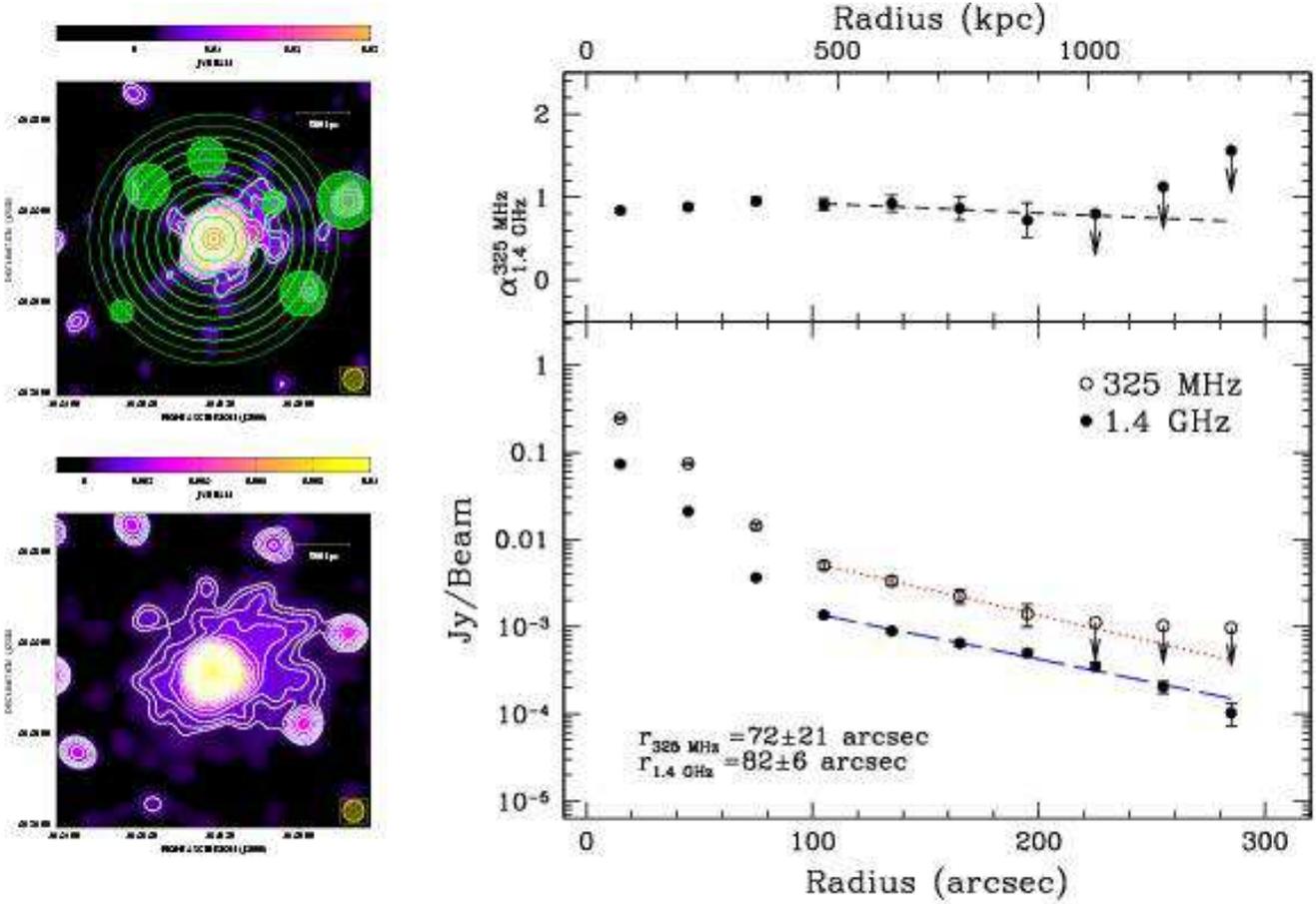}
\end{center}
\caption[]  { A2219. Left: Colors represent the images at 325\,MHz (top panel) and 1.4\,GHz (bottom panel). Contours represent the  325\,MHz and 1.4\,GHz images. Contours are spaced by $\sqrt{2}$ and start
 at 5.1 and 0.48  mJy/beam at 325\,MHz and 1.4\,GHz, respectively. The cross-hatched regions have 
been excluded from the statistics.
Right: Bottom panel shows the azimuthally averaged brightness profiles at 325\,MHz (open circles) and at 1.4\,GHz (solid circles). 
Arrows represent $3\sigma$ upper-limits. The lines represent the fit of an exponential profile to the data (see the text). In the top panel we plot the 
spectral index radial profile obtained from the brightness profiles.}  

\label{A2219_profi}
\end{figure*}

In Fig.~\ref{A2219_profi} (right), the brightness and the spectral profiles 
for the halo in A2219 are presented. The notation is the same as in Fig.\,\ref{A2744_profi}.
 As mentioned before, since in A2219 the radio emission in the center of the halo is dominated by 
the blend of three point sources, we focused only on the brightness and spectrum at
distances more than 500\,kpc from the cluster center.  The azimuthally averaged spectral index in the radio halo of A2219 is constant 
 within the error, with an average value of $\alpha\simeq 1$.

Due to the limited information on this radio halo, no attempt was made to derive constraints on 
the equipartition magnetic field profile for this cluster.

As for A2744, we compare results from radio observations to those from
the recent optical analysis based on photometric and spectroscopic
data for member galaxies (Boschin et al. 2004). The radio center 
coincides with the center of the galaxy distribution, i.e. the
position of the cD galaxy.  The western extension in the radio
emission finds an interesting possible correspondence in the western
extension of the 2D spatial distribution of bright galaxies (see
fig. 9 middle panel of Boschin et al. 2004) and in the East-West direction of the global
velocity gradient as recovered from the spectroscopic data.
 
\section{Discussion}
\label{reaccelmod}
 One of the main difficulties in explaining radio halos arises from the 
combination of their Mpc size and the short radiative lifetime of the 
relativistic electrons (about $10^{8}$ yrs): the diffusion time necessary
 to these electrons to cover such distances is much larger than their radiative
 lifetime.  Thus, it is required that either the electrons are 
re-accelerated ({\it primary models}) or continuously injected over the entire cluster 
volume by hadronic collisions ({\it secondary models}). Detailed studies of 
the spectral index in radio halos can provide important inputs to the above models.

For instance, secondary models assume that the relativistic electrons are continuously injected
 by cosmic ray protons colliding with thermal protons.
The cosmic ray protons are accelerated, with a power law energy spectrum, by cluster merger shocks and/or 
by galactic winds and then they are accumulated over cosmological epochs during the cluster formation 
(e.g. Berezinsky et al. 1997; Pfrommer et al. 2006 and references therein). The energy spectrum of the relativistic electrons produced 
 by the diffusion of the cosmic ray protons through the ICM is expected to be a smooth power law.
 In these models, variations in the halo's magnetic field strength do not produce variation in the 
radio spectral index. Thus, the constancy of the azimuthally average spectral index profile observed in 
A2744 and A2219 is in agreement with the expectation of secondary models. However, the patchy 
structure of the spectral index image in A2744 shows significant variations of the radio halo 
spectrum over distances as low as $\sim 200$ kpc. These variations cannot be explained by a 
featureless power law energy spectrum for the synchrotron electrons and indicate that more complex 
processes are at work in the ICM.

Primary models assume that the relativistic electrons are re-accelerated over the entire cluster.
The particle re-acceleration can be powered by the energy dissipated during cluster mergers.
The anti-correlation between spectral index and gas temperature shown 
in Fig.\,\ref{spix-T} supports the idea that a fraction of the 
gravitational energy, which is dissipated during major and minor
mergers in heating the thermal plasma, is converted into
re-acceleration of relativistic particles and amplification
of the magnetic field. In principle the re-acceleration of these particles may occur
either via shock acceleration or via turbulent re-acceleration;
both scenarios being qualitatively consistent.
However, the lack of a clear morphological connection
between the presence of shocks in the X-ray image and of synchrotron
emitting regions with flatter spectrum would apparently disfavor the shock hypothesis.

Complex spectral energy distributions for the synchrotron electrons 
are expected in primary models (Brunetti et al. 2001). Therefore, 
magnetic field variations within the radio halo can give rise
to spectral index patterns as those observed in the case of A2744.

Primary models assumes that electrons are re-accelerated
up to a maximum energy ($\gamma_{\rm max}$) which marks the balance
between acceleration efficiency and energy losses. Above $\gamma_{\rm
max}$ an exponential cut-off in the electron energy spectrum develops.  
The synchrotron emission extends up to a peak frequency $\nu_{\rm peak}
\propto \gamma_{\rm max}^2 B$, therefore the lower are $\gamma_{\rm max}$ and/or
$B$, the steeper is the synchrotron spectrum measured between two
fixed frequencies.  
In the re-acceleration scenario via the classical Fermi II mechanism,
the systematic energy gain of particles become significant after one
 acceleration time-scale, $\tau_{acc}$. If $\tau_{acc}$ does not depend on the particle energy, 
the maximum energy of the accelerated
particles essentially depends on the ratio between the acceleration
efficiency ($\propto\tau_{acc}^{-1}$ ) and the particle energy losses:

\begin{equation}
\gamma_{\rm max}
\propto
\frac{1}{(B^2 + B_{cmb}^2)\tau_{acc}}
\label{gmax}
\end{equation}

where $B_{cmb}=~3.2\cdot(1+z)^2\,\rm \mu G$ is the inverse Compton equivalent field\footnote{At the redshift of 
A2744 $B_{cmb}~=~5.5\, \rm \mu G$}.

In this case the peak frequency scales as:

\begin{equation}
\nu_{\rm peak} \propto {{ B}\over{\left( B^2 + B_{cmb}^2 \right)^2 \tau_{acc}^{2}}}
\label{nub}
\end{equation}

\begin{figure}[]
\begin{center}
\includegraphics[width=8cm]{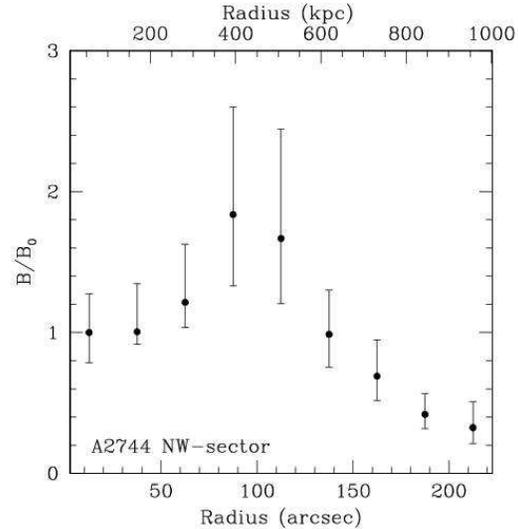}
\end{center}
\caption[]{Variation of the magnetic field strength in the case of constant acceleration efficiency 
in the NW quadrant of A2744. The magnetic field strength is normalized to its value at the cluster center.}
\label{A2744D2B}
\end{figure}

In the simplified hypothesis in which the acceleration efficiency is constant through the cluster and the magnetic field is smaller than the inverse Compton equivalent field,  $\nu_{\rm peak} \propto B$. Thus, we can infer the variations in 
 the magnetic field strength along those directions in which the observed spectral index shows the maximum changes, i.e. in the NW sector of A2744.
The shape of the energy spectrum of the synchrotron electrons depends on the adopted value for $\tau_{acc}$ and on the time for which the electrons are 
 accelerated (e.g., Ohno et al.~2002; Brunetti et al.~2004; Brunetti \& Blasi 2005; Cassano \& Brunetti 2005; Cho \& Lazarian 2006). 
Here we calculate the energy distribution of the re-accelerated electrons by assuming $\tau_{acc}= 10^8$yrs (which is appropriate for the emitting particles in 
radio halos) and by assuming that electrons are re-accelerated for a time-scale $\tau_H \sim 3 \tau_{acc}$ which corresponds to 0.3 Gyrs (in line with the 
age of radio halos, e.g. Hwang  2004).
In Fig.~\ref{A2744D2B} we report the radial behavior of the magnetic field strength obtained
by fitting the spectral index profile in the North West quadrant of A2744. In the approximation of constant acceleration efficiency, the  magnetic field strength in this quadrant of the cluster is constant, or slightly increasing, up to the core radius ($r_{c}\simeq 115$\arcsec) and it decreases by about a factor of two at the halo periphery.

\section{Summary}
We present new VLA images at 325\,MHz of the two clusters of galaxies   
A2744 and A2219, in which a wide diffuse emission was already detected at 1.4\,GHz. 
Combining the 325\,MHz and 1.4\,GHz data, we obtained the spectral index images 
 and the brightness radial profiles of the diffuse radio emission with a resolution of $\sim$ 1\arcmin~.

{\bf A2744:}  The radio emission of this cluster is characterized 
by the presence of a radio halo and a peripheral relic. The integrated spectral index between 1.4\,GHz and 325\,MHz 
is $\alpha~\simeq~$1 $\pm$~0.1 in the cluster and $\alpha~\simeq~ 1.1 \pm$~0.1 in the relic. 
The azimuthally averaged spectral index in A2744 is close to a value of $\alpha\simeq 1$ up to 1 Mpc from the cluster center.
However, the spectral index image is patchy, showing regions where the spectral index is significantly
different from the average. The observed spectral index variations range from a minimum of $\alpha \simeq 0.7\pm0.1$ to a
maximum $\alpha \simeq 1.5\pm0.2$.

From the comparison of the spectral index image and radial profiles with the optical data from Boschin et al. (2006)
and X-ray data (Kempner \& David 2004) it appears that the southern, flat spectrum, part of the radio halo coincides in 
projection with the high-velocity group of
 galaxies.  There is no evident correlation between the radio spectral index and the 
X-ray brightness substructures. However, the region of the NW group has a steep spectrum, while the the spectrum is flatter
 between the cluster center and the NW group. Moreover flat spectrum regions tend to have higher 
temperature. This result supports the idea that a fraction of the gravitational energy, which is dissipated during major and minor 
mergers in heating the thermal plasma, is converted into re-acceleration of relativistic particles and amplification of the magnetic 
field.

{\bf A2219:}  The radio emission in the central regions of the cluster is dominated by the blend of discrete sources. Therefore, only the outer regions of the radio halo can be studied. The radio spectrum has an average value of $\alpha\simeq 0.8$ in the central 
region and a constant profile with $\alpha\simeq 1$ in the radio halo. The limited sensitivity of the 325\,MHz image does not allow 
us to detect all the radio halo structure seen at 1.4\,GHz and therefore no constrains on the 
point-to-point variations of the spectral index have been obtained for this cluster.

\begin{acknowledgements}
We are grateful to Joshua C. Kempner for the {\it Chandra} images 
in electronic format he kindly provided us. 
We are thankful to Walter Boschin for the optical image he courteously offered us.
We thank Namir Kassim and Wendy Lane Peterson for advices and help. 
We thank the referee Rick Perley for his criticism and helpful suggestions, 
which pushed us to revise the data reduction. 
This work was partly supported by the Italian Ministry for University
and Research (MIUR) and by the Italian National Institute for Astrophysics (INAF). 
This research has made use of the
NASA/IPAC Extragalactic Data Base (NED) which is operated by the JPL, 
California Institute of Technology, under contract with the National 
Aeronautics and Space Administration.
\end{acknowledgements}

\end{document}